\documentclass[letterpaper,aps,superscriptaddress,showpacs,floatfix,twocolumn]{revtex4-1}

\usepackage{hyperref}
\usepackage{color}
\usepackage{graphicx}	
\usepackage{comment}
\graphicspath{%
  {./},%
}

\usepackage{xspace}	

\begin{document}

\title{Determining Transport Coefficients for a Microscopic Simulation of a Hadron Gas}

\author{Scott Pratt}
\affiliation{Department of Physics and Astronomy and National Superconducting Cyclotron Laboratory,
Michigan State University, East Lansing, Michigan 48824, USA}
\author{Alexander Baez}
\affiliation{Department of Physics and Astronomy and National Superconducting Cyclotron Laboratory,
Michigan State University, East Lansing, Michigan 48824, USA}
\affiliation{Department of Physics, University of South Florida, 4202 East Fowler Ave,
Tampa, Florida 33620-7100}
\author{Jane Kim}
\affiliation{Department of Physics and Astronomy and National Superconducting Cyclotron Laboratory,
Michigan State University, East Lansing, Michigan 48824, USA}

\date{\today}

\begin{abstract}

Quark-Gluon plasmas produced in relativistic heavy-ion collisions quickly expand and cool, entering a phase consisting of multiple interacting hadronic resonances just below the QCD deconfinement temperature, $T\sim 155$ MeV. Numerical microscopic simulations have emerged as the principal method for modeling the behavior of the hadronic stage of heavy-ion collisions, but the transport properties that characterize these simulations are not well understood.  Methods are presented here for extracting the shear viscosity, and two transport parameters that emerge in Israel-Stewart hydrodynamics. The analysis is based on studying how the stress-energy tensor responds to velocity gradients. Results agree with expectations based on Kubo relations.
\end{abstract}

\maketitle

The theory of strong interactions, QCD, predicts that at temperatures exceeding $T_c\simeq 155$ MeV, ordinary hadrons dissolve into their constituents creating a state of matter called the quark-gluon plasma (QGP). From first-principle lattice QCD studies, it is known that the transition from confined quarks and gluons (hadrons) to the deconfined quark-gluon plasma is not a true sharp phase transition but rather an analytic cross-over transition \cite{Aoki:2006we,Bhattacharya:2014ara}. Relativistic heavy-ion collisions, conducted at different collision energies at the Relativistic Heavy-Ion Collider (RHIC) and the Large Hadron Collider (LHC) aim at understanding the properties of QCD matter by creating quark-gluon plasmas at temperatures up to $\sim 450$ MeV, which then expand and cool to temperatures close to $T_c$ within a time of $\tau_0 \sim 5-10$ fm/c ($10^{-22}$ seconds). Below $T_c$, quarks and gluons confine into color-neutral objects (hadrons and hadron resonances) in a process called hadronization which is still not well understood. This hadron gas then continues to expand and cool, with the different hadrons scattering, decaying and merging for $\sim 10-20$ fm/$c$, before the remaining particles cease to interact and are eventually recorded by the experimental detectors. With the exception of direct photons and dileptons, all experimental information on the transport properties of hot QCD matter must be extracted from these final state hadrons. 

From the moment the nuclei first overlap, until the last particle interaction, systems created in heavy-ion collisions spend most of their lifetime in the hadronic gas phase. As such, inferring properties of the quark-gluon plasma phase from the measured particle spectrum with good precision naturally requires good understanding of the properties of the hadron resonance gas. While this is arguably the case at low temperatures where the system is well approximated as a gas of only pions \cite{Gavin:1985ph,Prakash:1993bt,Dobado:2001jf,Chen:2006iga,Chen:2007xe,Itakura:2007mx,Wiranata:2012br,Wiranata:2013oaa}, at temperatures close to $T_c$ the properties of this hadron resonance gas are not well known. The standard modeling procedure in heavy-ion collisions (which will also be used here) employs so-called hadron cascade codes to describe the hot hadron gas, which essentially consist of a kinetic theory simulation including all the hadronic resonance states found in the particle data book \cite{Olive:2016xmw}. Despite residual uncertainty in this model which arises from  poorly known cross sections in the temperature regime close to $T_c$, this approach is considered the best current available model. However, its transport properties close to $T_c$ are not well known. In previous works attempting to extract the value of shear viscosity over entropy density $\eta/s$ in a hadron gas, one study used the collision event generator URASIMA \cite{Muroya:2004pu} and two studies used a hadron cascade code (URQMD, Ref.~\cite{Bass:1998ca}), using different techniques \cite{Demir:2008tr,Song:2010aq}.
%
%
%
The present work is meant to improve our understanding of the hot hadron gas by providing procedures to extract transport properties from hadron cascades, and discuss findings in relation to our knowledge about transport in hot nuclear matter. One goal is to examine the stress-energy tensor to reliably extract effective viscosities and related shear transport coefficients, and to see whether the high values of the viscosity reported in \cite{Demir:2008tr} are reproduced. A second goal is to test whether the extracted viscosities are independent of the size of the imposed velocity gradients, which should be true for small velocity gradients, but has never been demonstrated for velocity gradients whose magnitudes are characteristic of those encountered in heavy-ion collisions.

\section{Methodology and Results}

In this study we consider the operational definition of shear viscosity from the Navier-Stokes (NS) equation,
\begin{equation}
\label{eq:NS}
T_{ij}=P\delta_{ij}-\zeta\nabla\cdot v-\eta(\partial_iv_j+\partial_jv_i-2\nabla\cdot v/3),
\end{equation}
where $P$ is the equilibrated pressure for the local energy density, and $\eta$ and $\zeta$ are the shear and bulk viscosities respectively. We will neglect the bulk viscosity for the moment and set $\zeta=0$. By analyzing a hadronic simulation with a chosen velocity gradient, one can observe the behavior of the stress-energy tensor, which  should be proportional to the shear viscosity multiplied by the velocity gradient.

On short time scales, the stress-energy tensor can be initialized to a broad range of values by adjusting the momenta of the simulated hadrons. For example, $T_{zz}$ can be initialized to zero if the particles are set with zero longitudinal velocity in the frame moving with the average fluid velocity. In Israel-Stewart hydrodynamics, see appendix, the deviation of the stress-energy tensor, $T_{ij}-P\delta_{ij}$, are dynamical variables that relax toward their Navier-Stokes values. Defining $\pi_{zz}\equiv T_{zz}-P$, the IS equations are
\begin{eqnarray}
\nonumber
\partial_t\pi_{ij}&=&-\frac{1}{\tau_\pi}\left(\pi_{ij}-\eta\omega_{ij}\right)
-\gamma_\pi\omega_{ik}\pi_{kj}-\kappa_\pi\nabla\cdot v\pi_{ij},\\
\label{eq:ISPhi}
\omega_{ij}&\equiv&\partial_iv_j+\partial_jv_i-(2/3)\delta_{ij}\nabla\cdot v.
\end{eqnarray}
The dimensionless coefficients $\kappa_\pi$ and $\gamma_\pi$ can depend on the temperature. For a massless gas with fixed cross sections, kinetic theory gives the coefficients as $\kappa_\pi=4/3$ and $\gamma_\pi=5/7$ \cite{Denicol:2012cn,Bazow:2013ifa,Denicol:2014tha,Bazow:2016yra}. A more general form for the coefficient $\kappa_\pi$ is derived in the appendix here,
\begin{eqnarray}
\label{eq:kappaalpha}
\kappa_\pi&=&s\frac{\partial_s\alpha}{\alpha},\\
\nonumber
\alpha^2&=&F_\pi s.
\end{eqnarray}
Here, $s$ is the entropy density and $F_\pi$ is the equal-time fluctuation of the traceless elements of the stress-energy tensor and is related to the viscosity and the relaxation time $\tau_\pi$ through a Kubo relation,
\begin{eqnarray}
\label{eq:kubo}
\eta&=&\frac{1}{2T}\int d^4r ~\langle T_{xy}(0)T_{xy}(r)\rangle\\
\nonumber
&\approx& \frac{\tau_\pi}{T}F_{\pi},\\
F_{\pi}&=&\sum_i(2S_i+1)\int \frac{d^3p}{(2\pi)^3}\frac{p_x^2p_y^2}{E_i({\bf p})^2}
e^{-E_i({\bf p})/T},\\
\nonumber
\frac{F_{\pi}}{T}&=&P-\frac{1}{15}\sum_i(2S_i+1)\int \frac{d^3p}{(2\pi)^3}\frac{p^4}{E_i^3}.
\end{eqnarray}
The expression for $F_\pi$ assumed the hadrons behave as a gas, i.e. the only correlations at equal times are those between a particle and itself. One can find the quantities above, including the entropy as a function of temperature, which allows one to also find $\kappa_\pi$. For the range of temperatures, $100<T<170$, $\kappa_\pi$ varied only by one \% from a value of 1.13.

During the course of the simulation, $\pi_{zz}$ and $F_\pi$ can be extracted from the momenta of the particles even when the system is not chemically equilibrated,
\begin{eqnarray}
\label{eq:Fpi}
P&=&\frac{1}{V}\sum_{i\in V}\frac{p_i^2}{3E_i}\\
\nonumber
\pi_{zz}&=&T_{zz}-P=\frac{1}{V}\sum_{i\in V}\frac{3p_{i,z}^2-p_i^2}{3E_i},\\
\nonumber
F_\pi&=&\frac{1}{15V}\sum_{i\in V} \frac{p_i^4}{E_i^2},\\
\frac{F_{\pi}}{T}&=&P-\frac{1}{15V}\sum_{i\in V} \frac{p_i^4}{E_i^3}.
\end{eqnarray}
Thus, given that $F_\pi$ and $F_\pi/T$ can be extracted from the cascade at any time, the Kubo formula provides a direct relation between the relaxation time and the viscosity, and the relaxation time is not a free parameter if given the viscosity. The last term in Eq. (\ref{eq:ISPhi}) has an unknown dimensionless coefficient $\gamma_\pi$. For kinetic theory of a relativistic gas, this coefficient is known to be $5/7$ \cite{Denicol:2012cn,Bazow:2013ifa}. However, for this case we treat it as an unknown, so we are left with two free parameters to describe the evolution of $\pi_{zz}$.  We then fit the behavior of the stress-energy tensor from a microscopic simulation performed in a controlled environment by varying both $\eta$ and $\gamma_\pi$. Since this also fixes the relaxation time, $\tau_\pi$, we can compare the relaxation time to the collision time and see whether $\tau_\pi\approx 2\tau_{\rm coll}$, which was found for the case studied in \cite{Denicol:2012cn}.

Here, we initialize our microscopic simulation, B3D \cite{Novak:2013bqa}, with the longitudinally boost-invariant geometry of Bjorken hydrodynamics \cite{Bjorken:1982qr}. In this environment the only velocity gradient in the fluid frame is $\partial_zv_z=1/\tau$, where $\tau=\sqrt{t^2-z^2}$ is the proper time, the time relative to when the collision began according to an observer moving with the fluid. All the matter is situated at $z=0$ at $\tau=0$, due to the large Lorentz contraction of the incoming nuclei, then it flows with a collective velocity $v_z=z/t$. In the frame of the matter, the behavior of all intrinsic quantities, including the stress-energy tensor, is chosen to depend only on the proper time, making the evolution effectively invariant to longitudinal boosts for observers beginning at $z=t=0$.

Particles were initialized at a proper time $\tau_0$ and temperature $T_0$, and with an initial anisotropy to the stress-energy tensor. The initial momenta were chosen according to the prescription in \cite{Pratt:2010jt}. This code was modified relative to previous versions to initially generate particles with a distribution of masses consistent with a thermal-weighted modified Lorentzian distribution. This change is described in the appendix. The spatial coordinates were spread uniformly in spatial rapidity, $\eta_s=\sinh^{-1}z/\tau$, so that the intrinsic properties depended only on $\tau$ and not $\eta$. The extent of the simulation in the transverse direction was confined to a radius of $R=40$ fm, with an extra buffer distance set up to ensure that the existence of the boundary could not affect the particles with radius $r<R$ during the finite time over which the cascade ran. The initial momentum distribution was modified to reproduce arbitrary initial $\pi_{zz}$ using the techniques of \cite{Pratt:2010jt}. The distribution of the particles longitudinally was confined to spatial rapidities $-\eta_{s,{\rm min}}<\eta<\eta_{s{\rm max}}$, with cyclic boundary conditions to longitudinal boosts. The scattering algorithm in B3D was written in terms of $\tau$ and $\eta$ and is explicitly invariant to longitudinal boosts. For instance, if collisions were ordered in the time $t$ rather than the proper time $\tau$, boost invariance could be violated. At high densities, cascade codes can differ from Boltzmann simulations due to the range of the interaction, $\sqrt{\sigma_\pi/\pi}$, extending to distances over which the collective velocity of the matter changes. This can be corrected by oversampling the particles by a factor $S$, then reducing the cross sections by the same factor to ensure the collision rates are not changed. For this calculation a sampling factor $S=4$ was applied for all calculations. Only the calculations at highest temperatures were noticeably affected by the oversampling. 

Runs were first performed for four initial temperature, $T=120, 135, 150$ and 165 MeV. To evaluate the behavior for a small velocity gradient, $\tau_0$ was set to $1000, 750, 500$ and 250 fm/$c$. Larger values of $\tau_0$ result in smaller viscous corrections, which then become increasingly difficult to analyze due to statistical noise. For these times the deviation $\pi_{zz}$ was $\sim$ one percent of the pressure. Smaller $\tau_0$ was used at higher temperature because the viscosities were smaller and using 1000 fm/$c$ would have led to values of $\pi_{zz}$ that were difficult to analyze due to statistical noise. The number of collisions and volumes were adjusted so that the typical number of analyzed particles would be on the order of $5\times 10^7$ particles.  At each time step, $P$, $F_\pi$  and $\pi_{zz}$ were determined by sampling the particles momenta using Eq. (\ref{eq:Fpi}). For each particle, the momenta in Eq. (\ref{eq:Fpi}) were evaluated in the local rest frame of the fluid as determined by the particle's position. The entropy density was taken by using the equilibrium entropy at $\tau_0$ scaled down by $\tau_0/\tau$.  These quantities then determine both $\alpha$ and $\kappa_\pi$ from Eq. (\ref{eq:kappaalpha}). For each of the four initial temperatures, three simulations were performed with different initial values of $\pi_{zz}(\tau_0)$. The values were chosen to correspond to NS values of $\pi_{zz}$ for $\eta/s=0.08,0.32$ and $1.28$.

In the Bjorken geometry, the IS equations become
\begin{eqnarray}
\label{eq:ISBj}
-\partial_\tau\pi_{zz}&=&-\frac{1}{\tau_\pi}\left(\pi_{zz}-\frac{4\eta}{3\tau}\right)-\frac{\kappa_\pi}{\tau}\pi_{zz}-\frac{4\gamma_\pi}{3\tau}\pi_{zz},\\
\partial_\tau\left(\frac{\pi_{zz}}{\alpha}\right)&=&-\frac{1}{\tau'_\pi}\left(\frac{\pi_{zz}}{\alpha}-\frac{4\eta'}{3\alpha\tau}\right),\\
\nonumber
\tau'_\pi&=&\tau_\pi/(1+4\gamma\tau_\pi/\tau),\\
\nonumber
\eta'&=&\eta/(1+4\gamma\tau_\pi/\tau).
\end{eqnarray}
The densities change modestly during the cascade, and thus both $\tau_\pi$ and $\gamma_\pi$ might change slightly during the evolution. We assume that the relaxation time $\tau_\pi$ scales inversely with the hadron density, whereas $\gamma_\pi$ is fixed. The two varied parameters are then the initial values of $\tau_\pi$ and $\gamma_\pi$. The differential equation in Eq. (\ref{eq:ISBj}) was then solved numerically, with the procedure repeated with different parameters varied until a best fit to the evolution was found. 

\begin{figure}
\centerline{\includegraphics[width=0.5\textwidth]{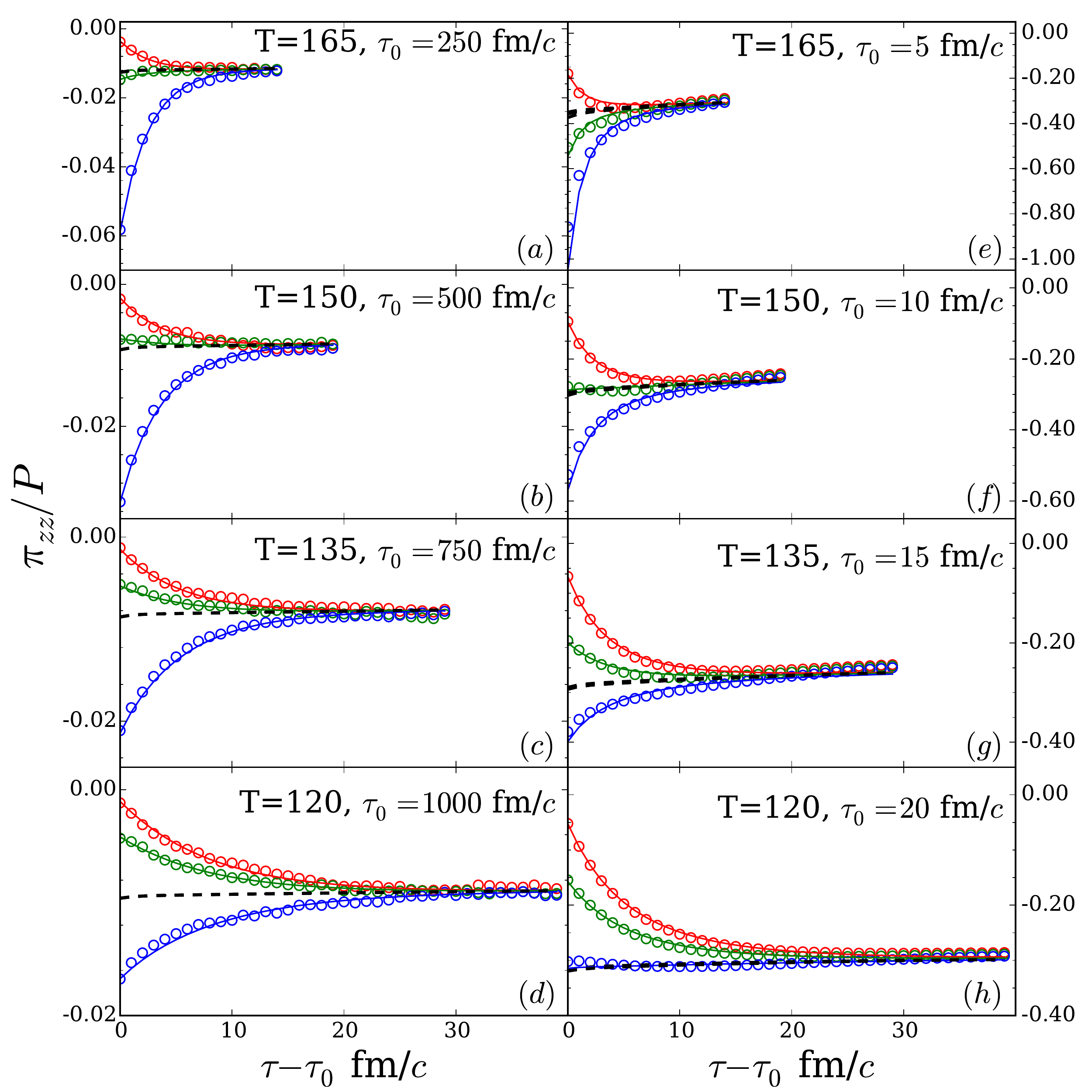}}
\caption{\label{fig:analfit}(color online) 
The evolution of the stress-energy tensor as extracted from the hadronic simulation (circles) are compared to Israel-Stewart forms (solid lines) to extract effective viscosities. Each separate line corresponds to a different initialization of the deviation of the stress-energy tensor, $\pi_{zz}$. In the left-side panels (a-d) small velocity gradients were imposed, equivalently large $\tau_0$, which should lead to true values of viscosity as $\tau_0$ is much larger than the collision time. Only one parameter, the effective viscosity $\eta^{\rm(eff)}$, was used to fit the three instances displayed for each panel, and the parameter $\gamma_\pi$ was set to zero. If the assumption of linearity for the Navier-Stokes equation is valid, the same values of $\eta^{\rm(eff)}$ would result from the right-side panels (e-h) which assume much larger velocity gradients, in the range of what might be produced in the environment of heavy-ion collisions.}
\end{figure}

\begin{figure}
\centerline{\includegraphics[width=0.5\textwidth]{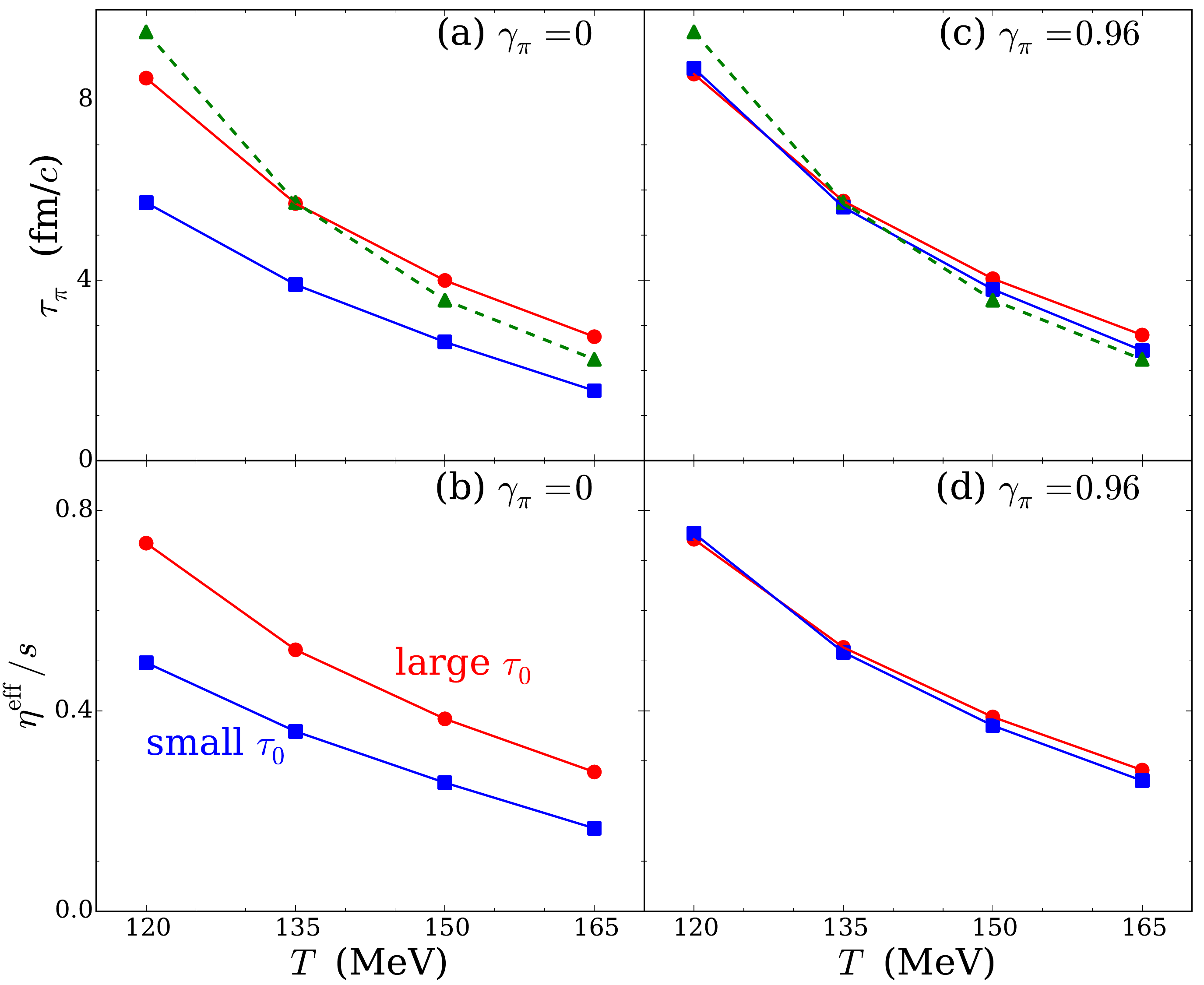}}
\caption{\label{fig:visc4}
(color online) The viscosity to entropy ratio extracted from calculations with small velocity gradients (red circles) differs from that calculated with large velocity gradients (blue circles), when one sets the viscous parameter $\gamma_\pi=0$ as illustrated in panel (b). By setting the value $\gamma_\pi=0.96$, the extracted viscosity becomes consistent for both velocity gradients as seen in panel (d). In the upper panels (a) and (c), the relaxation time $\tau_\pi$ is displayed, which mirrors the behavior of the viscosity. The extracted relaxation time in (c) consistently matches a value of two collision times, $2\tau_{\rm coll}$, which is represented by dashed lines.
}
\end{figure}

The left-side panels of Fig. \ref{fig:analfit} show the fit of $\pi_{zz}(\tau)$ for the four temperatures described above with large initial $\tau_0$. The IS equations reproduce both the relaxation of the stress-energy tensor toward the NS values, and the large $\tau$ behavior. For these large velocity gradients the effect of the $\gamma_\pi$ is negligible and one should extract the viscosity with confidence. The statistical noise in the extracted values are of the order of the size of the symbols in Fig. \ref{fig:analfit}, with the noise being correlated between different neighboring points due to the finite time required for relaxation. At each temperature, the fits are convincing for reproducing the evolution for three initial conditions. Figure \ref{fig:visc4} shows the extracted viscosities from these fits both with $\gamma_\pi=0.96$, and with $\gamma_\pi=0$. For the small velocity gradients, the extracted values of the viscosity and relaxation times are nearly independent of $\gamma_\pi$. Also, the extracted relaxation time $\tau_\pi$ closely matches twice the collision time as extracted from the cascade.

\begin{figure}
\centerline{\includegraphics[width=0.5\textwidth]{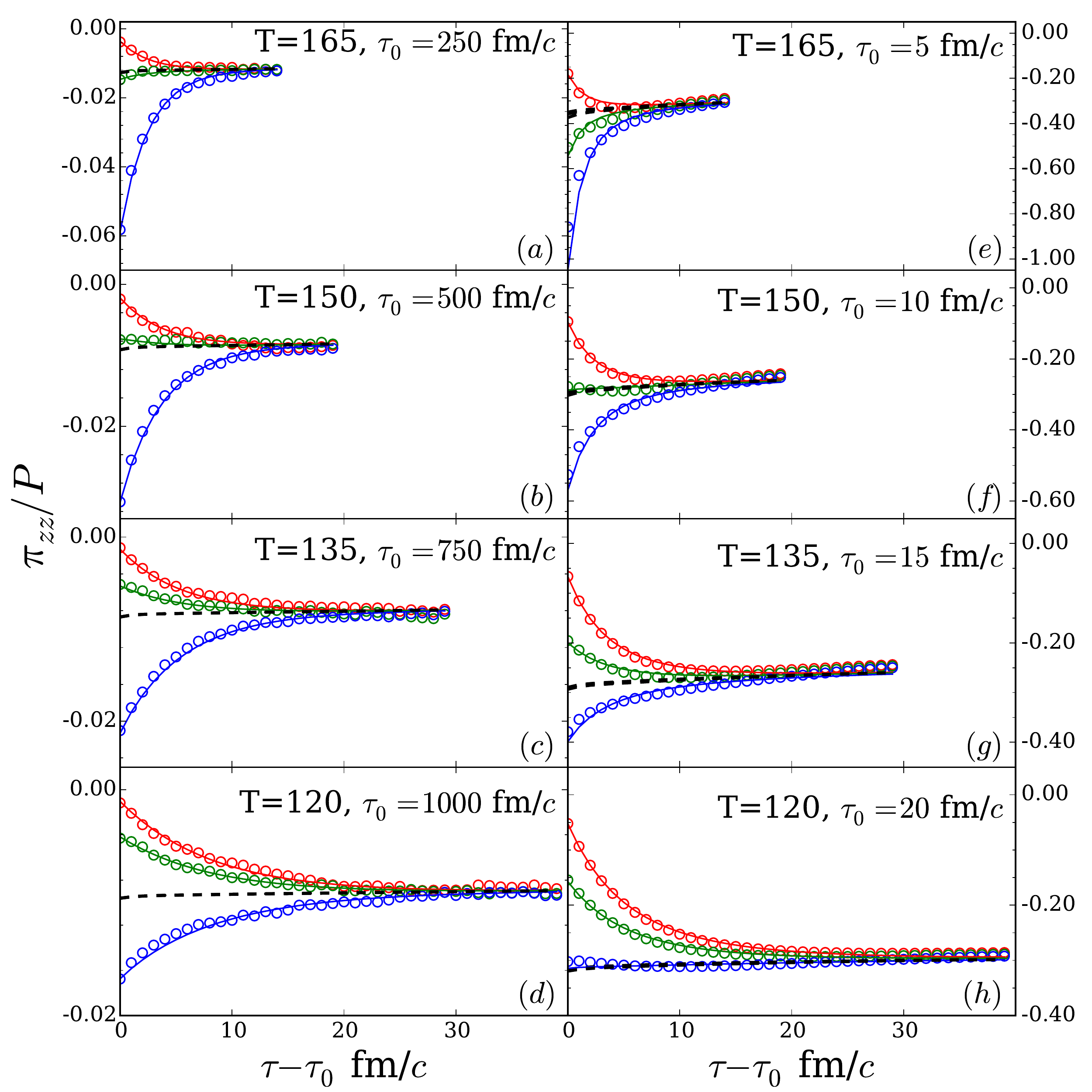}}
\caption{\label{fig:analfit_corrected}(color online) 
The same as Fig. \ref{fig:analfit}, only with $\gamma_\pi=0.96$. The fits to the same evolutions of the stress-energy tensor are practically indistinguishable from Fig. \ref{fig:analfit} where $\gamma_\pi$ was set to zero, but the extracted viscosities differed significantly for the cases with large velocity gradients, i.e. the small values of $\tau_0$ used in the right-side panels (e-h).}
\end{figure}

To determine $\gamma_\pi$ and to test the consistency of the approach, the procedure was repeated with large velocity gradients, by reducing $\tau_0$ by a factor of 50, to values in the neighborhood of what one would encounter in central collisions of heavy ions when traversing the hadron gas region. Again, the procedure was repeated with $\gamma_\pi=0$. Even though the fits to the evolution of $\pi_{zz}$, shown in the right-side panels of Fig. \ref{fig:analfit_corrected}, were again successful, the extracted viscosity and relaxation time were different than the values extracted with large $\tau_0$, or with small velocity gradients. Because $\tau_\pi$ was assumed to increase inversely with the hadron density, and because the hadron density behaves approximately inversely with $\tau$, the values for $\eta'$ and $\tau'_\pi$ in Eq. (\ref{eq:ISBj}) are effectively modified by a constant factor when $\gamma_\pi$ is introduced. Thus, the fits in Fig. \ref{fig:analfit} are indistinguishable with or without $\gamma_\pi$, but the values of the viscosities and relaxation times differ. Fig. \ref{fig:visc4} shows the extracted viscosities as a function of temperature assuming $\gamma_pi=0$ and assuming $\gamma_\pi=0.96$. For the value of 0.96, the viscosities and relaxation times are nearly the same for the large and small values of $\tau_0$. 

\begin{figure}
\centerline{\includegraphics[width=0.35\textwidth]{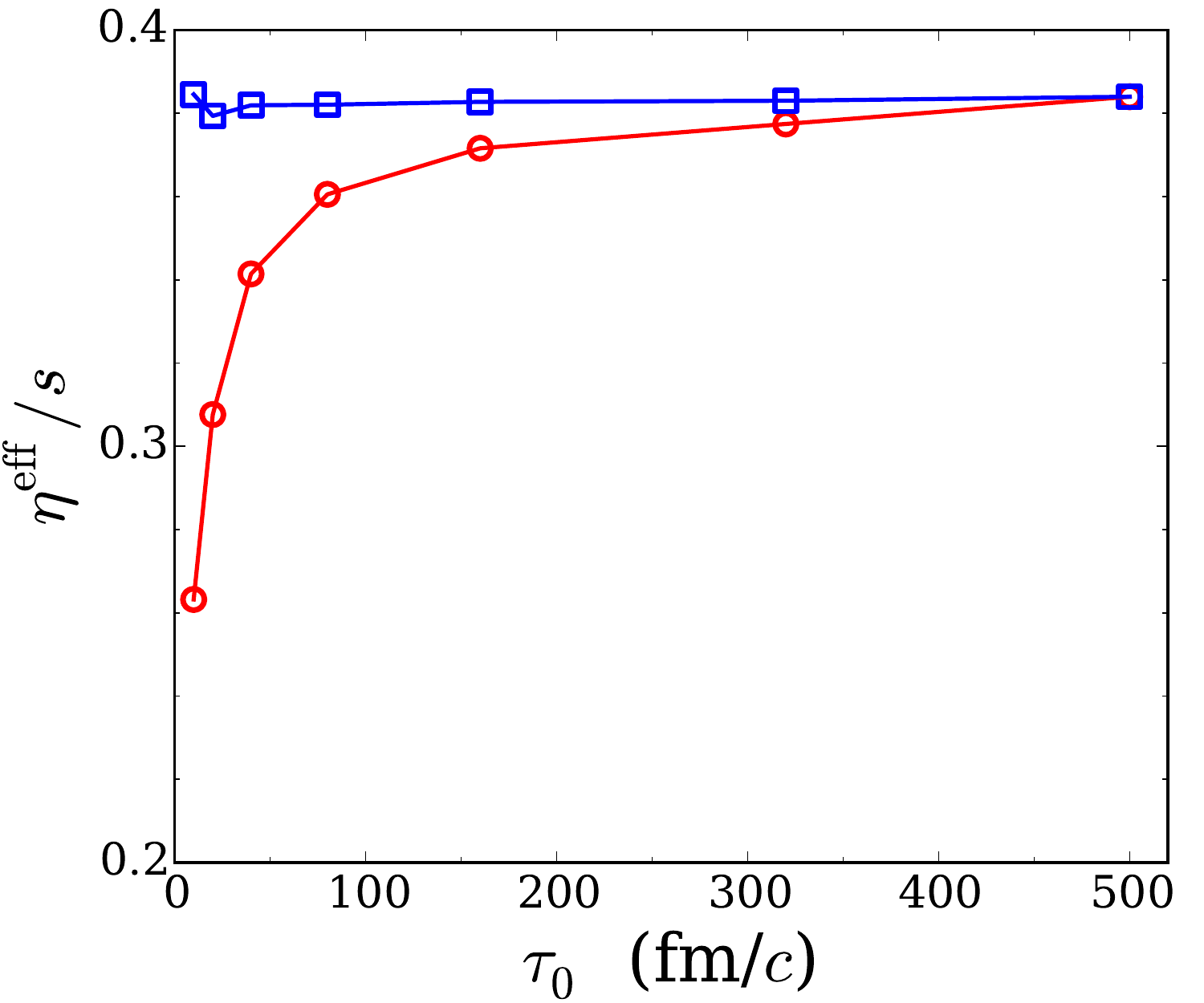}}
\caption{\label{fig:varioustau0}
(color online) The extracted viscosity to entropy ratio is shown for several runs which were all initialized with the same initial temperature, $T_0=150$ MeV, but had different initial velocity gradients, $1/\tau_0$. When setting $\gamma_\pi=0$ (red circles), the extracted viscosity was sensitive to the initial velocity gradient, but with $\gamma_\pi=0.96$ (blue squares), the results were independent of viscosity. This illustrates the importance of including the term in the Israel-Stewart equations for values of $\tau_0$ in the neighborhood of those encountered in heavy-ion physics, $\sim 5-10$ fm/$c$. 
}
\end{figure}

To more strongly demonstrate the effect of $\gamma_\pi\ne 0$, several initial times were studied for the initial temperature $T_0=150$ MeV. The extracted viscosities are shown in Fig. \ref{fig:varioustau0}, for both the case of $\gamma_\pi=0$ and $\gamma_\pi=0.96$. Effectively, the value of 0.96 was chosen by varying it until the viscosity in Fig. \ref{fig:varioustau0} was independent of $\tau_0$. 

\section{Conclusions}

Given that the analyses based on small velocity gradients resulted in shear effects where $\pi_{zz}/P$ is of the order of one percent, the extracted shear viscosities and relaxation times should be close to the correct values. However, the conditions of heavy-ion collisions can be quite different in that shear corrections can easily be tens of percent. To test whether this procedure works equally well for larger velocity gradients, the method was repeated with much smaller values of $\tau_0$, consistent with the scales inherent to where one samples the hadronic phase in heavy-ion collisions. By setting $\gamma_\pi=0.96$, the extracted values of the viscosity are independent of the velocity gradient, even for conditions where $\pi_{zz}/P$ is in the neighborhood of one third. This value is higher than the value of 5/7 found for a gas of massless particles with fixed cross-section \cite{Denicol:2012cn,Bazow:2013ifa}. Additionally, the IS coefficient $\kappa_\pi$ was determined for a hadron gas, and found to be near 1.13. This is somewhat smaller than the value, $4/3$, also determined in \cite{Denicol:2012cn,Bazow:2013ifa}. The analyses seem appear robust given the ability of a few parameters to reproduce the stress-energy tensor for a wide range of temperatures, velocity gradients, and initial anisotropies of the stress-energy tensor.

The bulk viscosity $\zeta$, and the bulk corrections to the pressure, were neglected here. Due to the fact that a hadron gas includes both highly relativistic particles, mainly the pions, and mostly non-relativistic particles, e.g. the baryons, the bulk viscosity is not negligible. However, stating the bulk viscosity requires comparing the stress-energy tensor to the equilibrated values, 
\begin{equation}
Pi=(T_{xx}+T_{yy}+T_{zz})/3-P.
\end{equation}
In addition to the deviations of the momentum dependence of the phase space distributions, the yields of the various hadrons rapidly lose equilibrium in the cooling system. Thus, one must decide whether to include the contribution to $\Pi$ from losing chemical equilibrium. Once the choice is made, it would be possible to follow the same procedure used here to determine $\zeta$. However, the complexity of the chemistry would be difficult to track given the large number of different hadronic resonances.

Determining the transport coefficients of the hadronic phase might motivate one to replace the microscopic simulation, the cascade, with a hydrodynamic description. However, that would be naive. In addition to the shear corrections, the bulk corrections are complicated due to the lack of chemical equilibrium. The loss of kinetic equilibrium is mainly driven by the various mass states losing thermal contact and having both different local temperatures, and different collective flows \cite{Pratt:1998gt,Sorge:1995pw}. Applying a hydrodynamic picture to account for the large number of species all moving with different collective flows and different local temperatures is untenable.

The value of the viscosity near the boundary of the hadronic phase is, nonetheless, important. If one believes that the true values of both $\eta$ and $\gamma_\pi$ are close to those of the cascade, it would guide choosing those parameters near the interface. Typically, one would apply this interface near $T=155$ MeV, the temperature where lattice calculations show that the hadron prescription begins to become unreasonable \cite{Bellwied:2015lba,Borsanyi:2011sw,Bazavov:2012jq}. The guidance taken from this analysis contrasts with that of \cite{Demir:2008tr}. In this study, one would expect $\eta/s\approx 0.3$ once the temperature nears 160 MeV, whereas the viscosities  in \cite{Demir:2008tr} had values $\eta/s\approx 1$ for all temperatures below 100 MeV. At low temperatures, the values found here are not far from those found in \cite{Wiranata:2012br,Wiranata:2013oaa} and \cite{Demir:2008tr}, but differ increasingly as the temperature rises from that point. The fact that the temperature dependence of both the viscosity and relaxation time were consistent with relaxation requiring two collisions at any temperature, leads credence to the values found here, whereas the the temperature independent values of $\eta/s$ in \cite{Demir:2008tr} are qualitatively at odds with expectations given falling collision times at higher density.

Another lesson from this analysis is the importance of the term with $\gamma_\pi$ in the IS equations, Eq. (\ref{eq:ISPhi}). Here, if one set $\gamma_\pi=0$, the extracted viscosity would have been understated by $\sim 30\%$ for calculations with velocity gradients having strengths characteristic of heavy-ion collisions. The statistical analysis of RHIC and LHC data in \cite{Pratt:2015zsa} indeed applied an IS hydrodynamic picture with $\gamma_\pi=0$. Because the physics of elliptic flow is driven by the stress-energy tensor, and if those anisotropies could be reproduced by calculations with higher viscosity if $\gamma_\pi$ was not set to zero, the extracted viscosities might be significantly larger, perhaps by several tens of percent, if $\gamma_\pi$ was set to reasonable value. If that were the case, the value of the viscosity for matter just above the interface temperature, might be found to be nearer to 3 to 4 times the AdS/CFT limit, $\eta/s=1/4\pi$, rather than the two to three times that was found in \cite{Pratt:2015zsa}. That higher value would be very much consistent with values extracted from the cascade here for temperatures near 160 MeV.

\section{Conclusions}

In this work, we have studied transport properties in the hot hadron gas by measuring the energy-momentum tensor in a hadron cascade simulation undergoing longitudinal expansion. We were able to extract a temperature dependent value of the shear viscosity over entropy density for temperature between 120 and 170 MeV. We furthermore investigated the possibility of extracting the bulk viscosity and second-order transport coefficients in the hot hadron gas. While partial chemical non-equilibrium effects seem to prohibit us from extracting $\zeta/s$, we were able to extract an estimate for the shear viscous relaxation time at $T=165$ MeV. Many aspects of our work can be improved in a straightforward manner and we expect our methodology to be useful in quantitative studies of transport in the hot hadron gas in the future.

\appendix*
\section{Constraining Israel Stewart Eq.s from Linear Response Theory}
Israel-Stewart (IS) equations differ from Navier-Stokes (NS) equations by treating the spatial components of the stress-energy tensor as dynamical variables. First, we use shorthand to define the spatial components of the stress-energy tensor and the velocity gradients,
\begin{eqnarray}
a_1&\equiv&\frac{1}{2}(T_{xx}-T_{yy}),\\
\nonumber
a_2&\equiv&\frac{1}{\sqrt{12}}\left(T_{xx}+T_{yy}-2T_{zz}\right),\\
\nonumber
a_3&\equiv& T_{xy},~~a_4\equiv T_{xz},~~a_5\equiv T_{yz},\\
\nonumber
b&\equiv&\frac{1}{3}\left(T_{xx}+T_{yy}+T_{zz}\right)-P,\\
\omega_1&\equiv&\partial_xv_x-\partial_yv_y,\\
\nonumber
\omega_2&\equiv&\frac{1}{\sqrt{3}}\left(\partial_xv_x+\partial_yv_y-2\partial_zv_z\right),\\
\nonumber
\omega_3&\equiv&\partial_xv_y+\partial_yv_x,~~\omega_4\equiv\partial_xv_z+\partial_zv_x,~~\omega_5\equiv\partial_yv_z+\partial_zv_y.
\end{eqnarray}
With these definitions the NS equations become
\begin{eqnarray}
a_i=-\eta\omega_i,~~b=-\zeta\nabla\cdot v,
\end{eqnarray}
where $\eta$ and $\zeta$ are the shear and bulk viscosities respectively, and the rate of work being done per unit volume is
\begin{equation}
\label{eq:work}
\sum_i a_i\omega_i+(P+b)\nabla\cdot v=\sum_{ij}T_{ij}\partial_iv_j.
\end{equation}
Assuming exponential decay of the correlations of the stress-energy tensor, the Kubo relations are
\begin{eqnarray}
\label{eq:kuboexp}
\eta&=&\frac{\tau_\pi}{T}F_\pi,~~\zeta=\frac{\tau_\Pi}{T}F_\Pi,\\
\nonumber
F_\pi&=&\int d^3r \left\langle a_j(0)a_j({\bf r})\right\rangle,
~~F_\Pi=\int d^3r \left\langle b(0)b({\bf r})\right\rangle.
\end{eqnarray}
The indices $j$ are not summed. The quantities $F_\pi$ and $F_\Pi$ describe the equal-time fluctuations of the tensor at equilibrium. For a gas equal-time correlations vanish except between a particle and itself. The shear correlations can then be calculated in either discrete or differential form
\begin{eqnarray}
F_\pi&=&\frac{1}{V}\sum_{i\in V} \frac{p_{i,x}^2p_{i,y}^2}{E_i^2},\\
\nonumber
&=&\sum_k (2S_k+1)\int\frac{d^3p}{(2\pi)^3}e^{-E_k(p)/T}\frac{p_x^2p_y^2}{E_k(p)^2},\\
\nonumber
&=&\frac{1}{30\pi^2}\sum_k(2S_k+1)\int\frac{p^6 dp}{E_k(p)^2}e^{-E_k(p)/T},\\
\nonumber
\frac{F_\pi}{T}&=&P-\frac{1}{30\pi^2}\sum_k(2S_k+1)\int\frac{p^6 dp}{E_k(p)^2}e^{-E_k(p)/T}.
\end{eqnarray}
The sum over $k$ describes a sum over all the species of the hadron gas with spins $S_k$. These quantities can be calculated as a function of the temperature for a thermalized system from the integral, and can be determined from the simulation in a non-thermalized system using the discrete sum. The bulk term $F_\Pi$ is small for a gas, and vanishes for either massless particles or for non-relativistic particles. We neglect it here. Given that $F_\pi$ is known, extracting the shear viscosity comes down to determining the relaxation time $\tau_\pi$. 

In the matter frame, the IS equations have the form
\begin{eqnarray}
\label{eq:ISoriginal}
\partial_t\pi_{ij}&=&-\frac{1}{\tau_\pi}\left\{\pi_{ij}-\eta\omega_{ij}\right\}-\kappa_\pi\pi_{ij}\nabla\cdot v\\
\nonumber
&&\hspace*{50pt}-\gamma_\pi \pi_{ik}\omega_{kj}.
\end{eqnarray}
Terms of higher order in the deviations, e.g. $\pi^2$ are neglected as this is considered an expansion in the inverse Reynolds number \cite{Denicol:2012cn}. Terms of higher order derivatives, e.g. $\omega^2$, involve non-local effects and should also be much smaller. In addition, such terms make the equations non-parabolic and can lead to super-luminar transport. At non-zero baryon density or if one considers bulk corrections, one must add additional terms that involve $b$ and fluctuations of the charge density.

For the immediate purpose, we neglect the 2nd-order term with $\gamma_\pi$. This term is known to be important \cite{Bazow:2016yra}, but is not needed for the following proof.  After setting $\gamma_\pi=0$ in Eq. (\ref{eq:ISoriginal}), the IS equations can be written as
\begin{eqnarray}
\label{eq:IS}
\partial_t\left(\frac{a_i}{\alpha}\right)&=&-\frac{1}{\tau_\pi}\left(\frac{a_i}{\alpha}-\frac{a_i^{\rm(NS)}}{\alpha}\right),
\end{eqnarray}
where $\alpha$ is some function of the temperature, related to $\kappa_\pi$ by
\begin{eqnarray}
\kappa_\pi\nabla\cdot v &=&-\frac{1}{\alpha}\partial_t\alpha\\
\kappa_\pi&=&\frac{s}{\alpha}\partial_s\alpha,
\end{eqnarray}
where the last step used conservation of entropy, $\partial_ts=-s\nabla\cdot v$.

In the next few lines we show that
\begin{equation}
\alpha^2=F_\pi s,
\end{equation}
by applying the constraint that the entropy must always rise, regardless of the signs of either the deviations of the stress-energy tensor, $a_i$ or $b$, or of the velocity gradients, $\omega_i$ or $\nabla\cdot v$.

The entropy of a volume $V$ for deviations of the anisotropy of the stress-energy tensor at fixed energy is
\begin{equation}
S=sV-\frac{1}{2}sV \sum_i\frac{a_i^2}{F_\pi s},
\end{equation}
ignoring the fluctuation of the bulk component for the moment.The appearance of $F_\pi$ ensures that the average $\langle a_i^2\rangle$, where the averaging is with the weight $e^{S}$ within a large volume, returns $F_\pi/V$. The rate of change of entropy production, including that due to changing $a_i$ is
\begin{eqnarray}
\label{eq:dSdt}
\frac{dS}{dt}&=&-\frac{V}{T}\sum_i a_i\omega_i-s\frac{a_i}{\sigma_\pi}\frac{d}{dt}\frac{a_i}{\sigma_\pi},\\
\nonumber
\sigma_\pi^2&\equiv&F_\pi s.
\end{eqnarray}
We have made use of the fact that the term $sV$ can be considered a constant here because entropy is nearly conserved, and the term is already being multiplied by $a_i^2$, which is already small. The first term accounts for the entropy rise of the entropy of $s$ due to the rise of the energy of the expansion in a system where $\nabla\cdot v=0$, and comes from the thermodynamic relation, Eq. (\ref{eq:work}). One can now insert the IS ansatz, Eq. (\ref{eq:IS}), into Eq. (\ref{eq:dSdt}) and find
\begin{eqnarray}
\frac{dS}{dt}&=&-V\sum_i\left\{
\frac{1}{T}a_i\omega_i-s\frac{a_i\alpha}{\sigma_\pi^2\tau_\pi}\left(\frac{a_i}{\alpha}+\frac{\eta\omega_i}{\alpha}\right)\right.\\
\nonumber
&&\left.+s\frac{a_i^2}{\sigma_\pi\alpha}\frac{d}{dt}\left(\frac{\alpha}{\sigma_\pi}\right)\right\}\\
\nonumber
&=&\frac{sV}{\tau_\pi\sigma_\pi^2}\sum_i a_i^2
-V\left(\frac{1}{T}-\frac{s\eta}{\sigma_\pi^2}\right)\sum_i a_i\omega_i,\\
\nonumber
&&+sV\frac{a_i^2}{\sigma_\pi\alpha}\frac{d}{dt}\left(\frac{\alpha}{\sigma_\pi}\right)
\end{eqnarray}
The Kubo relation, Eq. (\ref{eq:kuboexp}), ensures the middle term vanishes. The last term must vanish whether the ratio $\alpha/\sigma_\pi$ is rising or falling, to ensure that the entropy always increases. This implies that $\alpha$ and $\sigma_\pi$ are equal to within an arbitrary constant factor. Thus $\alpha^2=\sigma_\pi^2=F_0s$, and
\begin{equation}
\frac{dS}{dt}=\frac{V}{F_\pi \tau_\pi}\sum_ia_i^2=\frac{V}{T\eta}\sum_a a_i^2.
\end{equation}
If the stress-energy tensor relaxes to the NS value, $a_i=-\eta\omega_i$, entropy production approaches the usual value $dS/dt=(\eta V/T)\sum_i\omega_i^2$. The corresponding expressions for the bulk contributions involve $a_i\rightarrow b$ and $\omega_i\rightarrow \nabla\cdot v$.

Finally, we rewrite the IS equations in a form that can be compared to other work, in this case a massless gas expanding in a Bjorken geometry \cite{Denicol:2010xn}. Equation (\ref{eq:kubo}) can be expressed as
\begin{equation}
\frac{d}{dt}a_i=-\frac{1}{\tau_\pi}\left(a_i-a_i^{\rm(NS)}\right)+\frac{a_i}{\alpha}\frac{d}{dt}\alpha.
\end{equation}
For a massless gas $\alpha$ scales like the energy density and falls with the proper time as $\tau^{-4/3}$ so $\dot{\alpha}{\alpha}=-4/3\tau$, and
\begin{equation}
\frac{d}{dt}a_i=-\frac{1}{\tau_\pi}\left(a_i-a_i^{\rm(NS)}\right)-\frac{4}{3}\frac{a_i}{\tau}.
\end{equation}
This agrees with the corresponding expression in \cite{Denicol:2012cn,Bazow:2013ifa}. Finally, one can reintroduce the term proportional to $\omega\pi$ in Eq. (\ref{eq:ISoriginal}) and get
\begin{equation}
\frac{d}{dt}\pi_{zz}=-\frac{1}{\tau_\pi}\left(\pi_{zz}-\pi_{zz}^{\rm(NS)}\right)-\frac{4}{3}\frac{\pi_{zz}}{\tau}-\frac{4\gamma_\pi}{3}\frac{\pi_{zz}}{\tau}
\end{equation}
In \cite{Denicol:2012cn,Bazow:2013ifa}, the factor $\gamma_\pi$ was found to be 5/7 for a gas of massless particle with fixed relaxation time, whereas our best fit here gave $\gamma_\pi\approx 0.96$.

Although we ignore bulk corrections to the pressure, $b=(T_{xx}+T_{yy}+T_{zz})/3-P$, the same derivation above can be extended to include the bulk viscosity. In that case, the IS equations become
\begin{eqnarray}
\nonumber
\partial_\tau \pi_{ij}&=&-\frac{1}{\tau_\pi}\left(\pi_{ij}-\eta\omega_{ij}\right)-\kappa_\pi\pi_{ij}\nabla\cdot v
-\gamma_\pi\pi_{ik}\omega_{kj}-Jb\omega_{ij},\\
\partial_\tau b&=&-\frac{1}{\tau_b}\left(b-\zeta\nabla\cdot v\right)-\kappa_b b\nabla\cdot v
-K{\rm Tr}~\pi\omega.
\end{eqnarray}
The same entropy arguments used for the evolution of $\pi_{ij}$, can be applied above to give
\begin{eqnarray}
F_\Pi&=&\int d^3r b(0)b(r),\\
\nonumber
\zeta&=&\tau_bF_\Pi/T,\\
\nonumber
\beta^2&=&F_\Pi s,\\
\nonumber
\kappa_\Pi&=&s\frac{\partial_s\beta}{\beta},\\
\nonumber
JF_\Pi&=&KF_\pi.
\end{eqnarray}
Thus, given $F_\Pi$, there are two additional independent parameters if one includes the bulk viscosity: the bulk viscosity $\zeta$, and either $J$ or $K$.

\section{Generating Initial Thermal Masses}
In order to instantiate the cascade B3D, particles are generated consistently with their distributions in a thermalized system. However, the previous version of B3D had a shortcoming, in that particles were initially assigned their pole mass, rather than a distribution consistent with the spectral function. In this study a mass distribution, or spectral function $S(M)$, was implemented with the form of a modified Lorentzian,
\begin{eqnarray}
\label{eq:spectral}
\frac{dN}{dM}&=&S(M)=\frac{2}{\Gamma_R\pi}\frac{(\Gamma/2)^2}{(M-M_R)^2+(\Gamma/2)^2},\\
\nonumber
\Gamma&=&\Gamma_R\left(\frac{2k^2}{k_R^2+k^2}\right)^{\alpha}.
\end{eqnarray}
Here, $k$ is the momentum of one of the particles in the center-of-mass frame where their momenta are $\vec{k}$ and $-\vec{k}$. The parameter $\alpha$ was chosen to be 1/2, but a more realistic model could have chosen different values depending on the angular momentum of the resonance. The momentum required to reach the resonance energy, $M_R$, is $k_R$. The cross section for creating a resonance was chosen to have the form,
\begin{equation}
\sigma=\frac{4\pi}{k^2}\frac{(\Gamma/2)^2}{(M-M_R)^2+(\Gamma/2)^2},
\end{equation}
and the lifetime of the resonance was chosen to be $\Gamma_R$, independent of the off-shellness. One can calculate the fraction of time spent in the resonance for particles of that energy in a volume $V$ by multiplying the collision rate and the lifetime,
\begin{equation}
f_R=\frac{\sigma v_{\rm rel}}{\Gamma_R V},
\end{equation}
where $v_{\rm rel}=|v_1|+|v_2|$ is the relative velocity. According to the Ergodic theorem \cite{Danielewicz:1995ay}, this should be the ratio of the spectral function to the density of states in the continuum,
\begin{eqnarray}
\frac{S(M)}{dN_{\rm cont}/dM}=\frac{S(M)}{(2\pi)^{-3}4\pi k^2/(dM/dk)}.
\end{eqnarray}
Since $dM/dK=dE_1/dk+dE_2/dk=|v_1|+|v_2|$, one can see that this ratio indeed equals $f_R$, and the method is ergodically consistent. If a system is created in a box with this mass distribution, that mass distribution should remain the same throughout time due this consistency. However, due to the fact that the inverse processes for three-body decays are missing from B3D, and because for multichannel systems the mass distribution is calculated using only the principal decay channel, the system properties might relax somewhat away from those of the system immediately after creation.

\begin{figure}
\centerline{\includegraphics[width=0.35\textwidth]{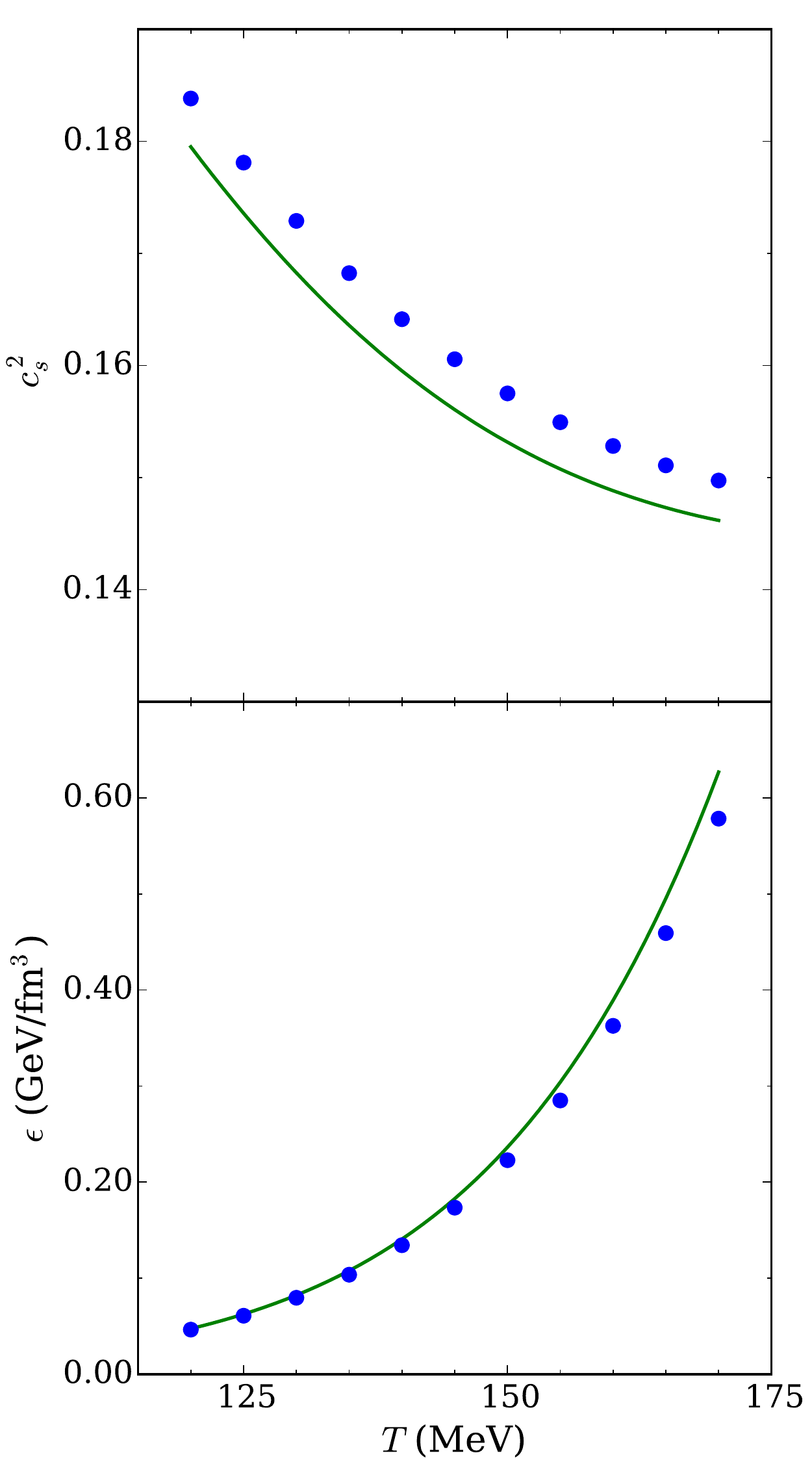}}
\caption{\label{fig:eos_alpha}(color online)
Compared to using the pole mass (green line) to calculate densities of resonances, the altered equation of state from the spectral function in Eq. (\ref{eq:spectral}) (blue points) is somewhat stiffer, with energy densities lowered by a few percent and the speed of sound raised by a few percent.
}
\end{figure}

Given the spectral function, the densities of each species are found by integrating over the densities of possible masses weighted by $S(M)$. Once the decision has been made to create a given species, masses are first chosen according to $S(M)$. This tends to skew the choice toward masses above $M_r$ due to the momentum dependence of the width, unless $\alpha=0$, and due to the fact that the lower end of masses is cut off due to kinematics, $M$ must be greater than the sum of the two decaying masses $m_1+m_2$. The mass $M$ is then kept or rejected proportional to the density of particles with that mass. This thermal weighting then skews the mass downwards. The average masses can then be either greater or less than $M_r$ depending on how much larger $M$ is relative to $m1$ and $m_2$, the choice of $\alpha$, and the temperature. Figure \ref{fig:eos_alpha} shows how the energy density and speed of sound are changed by this procedure compared to generating all particles with their pole mass. Energy densities are lowered by $\sim$ 3\% at higher temperatures, and the speed of sound is raised by a few percent. 

\begin{acknowledgments}
This work was supported by the Department of Energy Office of Science, awards No. DE-SC0008132 and DE-FG02-03ER41259. Alexander Baez acknowledges support from the National Science Foundation, Research Experience for Undergraduates, Division of Physics, award no. 1559866, and Jane Kim was supported by the Director's Research Scholar Program at the National Superconducting Cyclotron Laboratory, funded by the National Science Foundation through award no. 1102511. The authors thank Paul Romatschke for valuable discussions and helping with writing the manuscript.
\end{acknowledgments}

\bibliographystyle{apsrev} \bibliography{cascade}

\begin{thebibliography}{30}
\expandafter\ifx\csname natexlab\endcsname\relax\def\natexlab#1{#1}\fi
\expandafter\ifx\csname bibnamefont\endcsname\relax
  \def\bibnamefont#1{#1}\fi
\expandafter\ifx\csname bibfnamefont\endcsname\relax
  \def\bibfnamefont#1{#1}\fi
\expandafter\ifx\csname citenamefont\endcsname\relax
  \def\citenamefont#1{#1}\fi
\expandafter\ifx\csname url\endcsname\relax
  \def\url#1{\texttt{#1}}\fi
\expandafter\ifx\csname urlprefix\endcsname\relax\def\urlprefix{URL }\fi
\providecommand{\bibinfo}[2]{#2}
\providecommand{\eprint}[2][]{\url{#2}}

\bibitem[{\citenamefont{Aoki et~al.}(2006)\citenamefont{Aoki, Endrodi, Fodor,
  Katz, and Szabo}}]{Aoki:2006we}
\bibinfo{author}{\bibfnamefont{Y.}~\bibnamefont{Aoki}},
  \bibinfo{author}{\bibfnamefont{G.}~\bibnamefont{Endrodi}},
  \bibinfo{author}{\bibfnamefont{Z.}~\bibnamefont{Fodor}},
  \bibinfo{author}{\bibfnamefont{S.}~\bibnamefont{Katz}}, \bibnamefont{and}
  \bibinfo{author}{\bibfnamefont{K.}~\bibnamefont{Szabo}},
  \bibinfo{journal}{Nature} \textbf{\bibinfo{volume}{443}},
  \bibinfo{pages}{675} (\bibinfo{year}{2006}), \eprint{hep-lat/0611014}.

\bibitem[{\citenamefont{Bhattacharya et~al.}(2014)\citenamefont{Bhattacharya,
  Buchoff, Christ, Ding, Gupta et~al.}}]{Bhattacharya:2014ara}
\bibinfo{author}{\bibfnamefont{T.}~\bibnamefont{Bhattacharya}},
  \bibinfo{author}{\bibfnamefont{M.~I.} \bibnamefont{Buchoff}},
  \bibinfo{author}{\bibfnamefont{N.~H.} \bibnamefont{Christ}},
  \bibinfo{author}{\bibfnamefont{H.~T.} \bibnamefont{Ding}},
  \bibinfo{author}{\bibfnamefont{R.}~\bibnamefont{Gupta}}, \bibnamefont{et~al.}
  (\bibinfo{year}{2014}), \eprint{1402.5175}.

\bibitem[{\citenamefont{Gavin}(1985)}]{Gavin:1985ph}
\bibinfo{author}{\bibfnamefont{S.}~\bibnamefont{Gavin}},
  \bibinfo{journal}{Nucl.Phys.} \textbf{\bibinfo{volume}{A435}},
  \bibinfo{pages}{826} (\bibinfo{year}{1985}).

\bibitem[{\citenamefont{Prakash et~al.}(1993)\citenamefont{Prakash, Prakash,
  Venugopalan, and Welke}}]{Prakash:1993bt}
\bibinfo{author}{\bibfnamefont{M.}~\bibnamefont{Prakash}},
  \bibinfo{author}{\bibfnamefont{M.}~\bibnamefont{Prakash}},
  \bibinfo{author}{\bibfnamefont{R.}~\bibnamefont{Venugopalan}},
  \bibnamefont{and} \bibinfo{author}{\bibfnamefont{G.}~\bibnamefont{Welke}},
  \bibinfo{journal}{Phys.Rept.} \textbf{\bibinfo{volume}{227}},
  \bibinfo{pages}{321} (\bibinfo{year}{1993}).

\bibitem[{\citenamefont{Dobado and Santalla}(2002)}]{Dobado:2001jf}
\bibinfo{author}{\bibfnamefont{A.}~\bibnamefont{Dobado}} \bibnamefont{and}
  \bibinfo{author}{\bibfnamefont{S.~N.} \bibnamefont{Santalla}},
  \bibinfo{journal}{Phys.Rev.} \textbf{\bibinfo{volume}{D65}},
  \bibinfo{pages}{096011} (\bibinfo{year}{2002}), \eprint{hep-ph/0112299}.

\bibitem[{\citenamefont{Chen and Nakano}(2007)}]{Chen:2006iga}
\bibinfo{author}{\bibfnamefont{J.-W.} \bibnamefont{Chen}} \bibnamefont{and}
  \bibinfo{author}{\bibfnamefont{E.}~\bibnamefont{Nakano}},
  \bibinfo{journal}{Phys.Lett.} \textbf{\bibinfo{volume}{B647}},
  \bibinfo{pages}{371} (\bibinfo{year}{2007}), \eprint{hep-ph/0604138}.

\bibitem[{\citenamefont{Chen et~al.}(2007)\citenamefont{Chen, Li, Liu, and
  Nakano}}]{Chen:2007xe}
\bibinfo{author}{\bibfnamefont{J.-W.} \bibnamefont{Chen}},
  \bibinfo{author}{\bibfnamefont{Y.-H.} \bibnamefont{Li}},
  \bibinfo{author}{\bibfnamefont{Y.-F.} \bibnamefont{Liu}}, \bibnamefont{and}
  \bibinfo{author}{\bibfnamefont{E.}~\bibnamefont{Nakano}},
  \bibinfo{journal}{Phys.Rev.} \textbf{\bibinfo{volume}{D76}},
  \bibinfo{pages}{114011} (\bibinfo{year}{2007}), \eprint{hep-ph/0703230}.

\bibitem[{\citenamefont{Itakura et~al.}(2008)\citenamefont{Itakura, Morimatsu,
  and Otomo}}]{Itakura:2007mx}
\bibinfo{author}{\bibfnamefont{K.}~\bibnamefont{Itakura}},
  \bibinfo{author}{\bibfnamefont{O.}~\bibnamefont{Morimatsu}},
  \bibnamefont{and} \bibinfo{author}{\bibfnamefont{H.}~\bibnamefont{Otomo}},
  \bibinfo{journal}{Phys.Rev.} \textbf{\bibinfo{volume}{D77}},
  \bibinfo{pages}{014014} (\bibinfo{year}{2008}), \eprint{0711.1034}.

\bibitem[{\citenamefont{Wiranata and Prakash}(2012)}]{Wiranata:2012br}
\bibinfo{author}{\bibfnamefont{A.}~\bibnamefont{Wiranata}} \bibnamefont{and}
  \bibinfo{author}{\bibfnamefont{M.}~\bibnamefont{Prakash}},
  \bibinfo{journal}{Phys.Rev.} \textbf{\bibinfo{volume}{C85}},
  \bibinfo{pages}{054908} (\bibinfo{year}{2012}), \eprint{1203.0281}.

\bibitem[{\citenamefont{Wiranata et~al.}(2013)\citenamefont{Wiranata, Koch,
  Prakash, and Wang}}]{Wiranata:2013oaa}
\bibinfo{author}{\bibfnamefont{A.}~\bibnamefont{Wiranata}},
  \bibinfo{author}{\bibfnamefont{V.}~\bibnamefont{Koch}},
  \bibinfo{author}{\bibfnamefont{M.}~\bibnamefont{Prakash}}, \bibnamefont{and}
  \bibinfo{author}{\bibfnamefont{X.~N.} \bibnamefont{Wang}},
  \bibinfo{journal}{Phys.Rev.} \textbf{\bibinfo{volume}{C88}},
  \bibinfo{pages}{044917} (\bibinfo{year}{2013}), \eprint{1307.4681}.

\bibitem[{\citenamefont{Olive}(2016)}]{Olive:2016xmw}
\bibinfo{author}{\bibfnamefont{K.~A.} \bibnamefont{Olive}},
  \bibinfo{journal}{Chin. Phys.} \textbf{\bibinfo{volume}{C40}},
  \bibinfo{pages}{100001} (\bibinfo{year}{2016}).

\bibitem[{\citenamefont{Muroya and Sasaki}(2005)}]{Muroya:2004pu}
\bibinfo{author}{\bibfnamefont{S.}~\bibnamefont{Muroya}} \bibnamefont{and}
  \bibinfo{author}{\bibfnamefont{N.}~\bibnamefont{Sasaki}},
  \bibinfo{journal}{Prog. Theor. Phys.} \textbf{\bibinfo{volume}{113}},
  \bibinfo{pages}{457} (\bibinfo{year}{2005}), \eprint{nucl-th/0408055}.

\bibitem[{\citenamefont{Bass et~al.}(1998)\citenamefont{Bass, Belkacem,
  Bleicher, Brandstetter, Bravina et~al.}}]{Bass:1998ca}
\bibinfo{author}{\bibfnamefont{S.}~\bibnamefont{Bass}},
  \bibinfo{author}{\bibfnamefont{M.}~\bibnamefont{Belkacem}},
  \bibinfo{author}{\bibfnamefont{M.}~\bibnamefont{Bleicher}},
  \bibinfo{author}{\bibfnamefont{M.}~\bibnamefont{Brandstetter}},
  \bibinfo{author}{\bibfnamefont{L.}~\bibnamefont{Bravina}},
  \bibnamefont{et~al.}, \bibinfo{journal}{Prog.Part.Nucl.Phys.}
  \textbf{\bibinfo{volume}{41}}, \bibinfo{pages}{255} (\bibinfo{year}{1998}),
  \eprint{nucl-th/9803035}.

\bibitem[{\citenamefont{Demir and Bass}(2009)}]{Demir:2008tr}
\bibinfo{author}{\bibfnamefont{N.}~\bibnamefont{Demir}} \bibnamefont{and}
  \bibinfo{author}{\bibfnamefont{S.~A.} \bibnamefont{Bass}},
  \bibinfo{journal}{Phys.Rev.Lett.} \textbf{\bibinfo{volume}{102}},
  \bibinfo{pages}{172302} (\bibinfo{year}{2009}), \eprint{0812.2422}.

\bibitem[{\citenamefont{Song et~al.}(2011)\citenamefont{Song, Bass, and
  Heinz}}]{Song:2010aq}
\bibinfo{author}{\bibfnamefont{H.}~\bibnamefont{Song}},
  \bibinfo{author}{\bibfnamefont{S.~A.} \bibnamefont{Bass}}, \bibnamefont{and}
  \bibinfo{author}{\bibfnamefont{U.}~\bibnamefont{Heinz}},
  \bibinfo{journal}{Phys.Rev.} \textbf{\bibinfo{volume}{C83}},
  \bibinfo{pages}{024912} (\bibinfo{year}{2011}), \eprint{1012.0555}.

\bibitem[{\citenamefont{Denicol et~al.}(2012)\citenamefont{Denicol, Niemi,
  Molnar, and Rischke}}]{Denicol:2012cn}
\bibinfo{author}{\bibfnamefont{G.~S.} \bibnamefont{Denicol}},
  \bibinfo{author}{\bibfnamefont{H.}~\bibnamefont{Niemi}},
  \bibinfo{author}{\bibfnamefont{E.}~\bibnamefont{Molnar}}, \bibnamefont{and}
  \bibinfo{author}{\bibfnamefont{D.~H.} \bibnamefont{Rischke}},
  \bibinfo{journal}{Phys. Rev.} \textbf{\bibinfo{volume}{D85}},
  \bibinfo{pages}{114047} (\bibinfo{year}{2012}), \bibinfo{note}{[Erratum:
  Phys. Rev.D91,no.3,039902(2015)]}, \eprint{1202.4551}.

\bibitem[{\citenamefont{Bazow et~al.}(2014)\citenamefont{Bazow, Heinz, and
  Strickland}}]{Bazow:2013ifa}
\bibinfo{author}{\bibfnamefont{D.}~\bibnamefont{Bazow}},
  \bibinfo{author}{\bibfnamefont{U.~W.} \bibnamefont{Heinz}}, \bibnamefont{and}
  \bibinfo{author}{\bibfnamefont{M.}~\bibnamefont{Strickland}},
  \bibinfo{journal}{Phys. Rev.} \textbf{\bibinfo{volume}{C90}},
  \bibinfo{pages}{054910} (\bibinfo{year}{2014}), \eprint{1311.6720}.

\bibitem[{\citenamefont{Denicol et~al.}(2014)\citenamefont{Denicol, Heinz,
  Martinez, Noronha, and Strickland}}]{Denicol:2014tha}
\bibinfo{author}{\bibfnamefont{G.~S.} \bibnamefont{Denicol}},
  \bibinfo{author}{\bibfnamefont{U.~W.} \bibnamefont{Heinz}},
  \bibinfo{author}{\bibfnamefont{M.}~\bibnamefont{Martinez}},
  \bibinfo{author}{\bibfnamefont{J.}~\bibnamefont{Noronha}}, \bibnamefont{and}
  \bibinfo{author}{\bibfnamefont{M.}~\bibnamefont{Strickland}},
  \bibinfo{journal}{Phys. Rev.} \textbf{\bibinfo{volume}{D90}},
  \bibinfo{pages}{125026} (\bibinfo{year}{2014}), \eprint{1408.7048}.

\bibitem[{\citenamefont{Bazow et~al.}(2016)\citenamefont{Bazow, Heinz, and
  Strickland}}]{Bazow:2016yra}
\bibinfo{author}{\bibfnamefont{D.}~\bibnamefont{Bazow}},
  \bibinfo{author}{\bibfnamefont{U.~W.} \bibnamefont{Heinz}}, \bibnamefont{and}
  \bibinfo{author}{\bibfnamefont{M.}~\bibnamefont{Strickland}}
  (\bibinfo{year}{2016}), \eprint{1608.06577}.

\bibitem[{\citenamefont{Novak et~al.}(2013)\citenamefont{Novak, Novak, Pratt,
  Coleman-Smith, and Wolpert}}]{Novak:2013bqa}
\bibinfo{author}{\bibfnamefont{J.}~\bibnamefont{Novak}},
  \bibinfo{author}{\bibfnamefont{K.}~\bibnamefont{Novak}},
  \bibinfo{author}{\bibfnamefont{S.}~\bibnamefont{Pratt}},
  \bibinfo{author}{\bibfnamefont{C.}~\bibnamefont{Coleman-Smith}},
  \bibnamefont{and} \bibinfo{author}{\bibfnamefont{R.}~\bibnamefont{Wolpert}}
  (\bibinfo{year}{2013}), \eprint{1303.5769}.

\bibitem[{\citenamefont{Bjorken}(1983)}]{Bjorken:1982qr}
\bibinfo{author}{\bibfnamefont{J.~D.} \bibnamefont{Bjorken}},
  \bibinfo{journal}{Phys. Rev.} \textbf{\bibinfo{volume}{D27}},
  \bibinfo{pages}{140} (\bibinfo{year}{1983}).

\bibitem[{\citenamefont{Pratt and Torrieri}(2010)}]{Pratt:2010jt}
\bibinfo{author}{\bibfnamefont{S.}~\bibnamefont{Pratt}} \bibnamefont{and}
  \bibinfo{author}{\bibfnamefont{G.}~\bibnamefont{Torrieri}},
  \bibinfo{journal}{Phys.Rev.} \textbf{\bibinfo{volume}{C82}},
  \bibinfo{pages}{044901} (\bibinfo{year}{2010}), \eprint{1003.0413}.

\bibitem[{\citenamefont{Pratt and Murray}(1998)}]{Pratt:1998gt}
\bibinfo{author}{\bibfnamefont{S.}~\bibnamefont{Pratt}} \bibnamefont{and}
  \bibinfo{author}{\bibfnamefont{J.}~\bibnamefont{Murray}},
  \bibinfo{journal}{Phys. Rev.} \textbf{\bibinfo{volume}{C57}},
  \bibinfo{pages}{1907} (\bibinfo{year}{1998}).

\bibitem[{\citenamefont{Sorge}(1996)}]{Sorge:1995pw}
\bibinfo{author}{\bibfnamefont{H.}~\bibnamefont{Sorge}},
  \bibinfo{journal}{Phys. Lett.} \textbf{\bibinfo{volume}{B373}},
  \bibinfo{pages}{16} (\bibinfo{year}{1996}), \eprint{nucl-th/9510056}.

\bibitem[{\citenamefont{Bellwied et~al.}(2015)\citenamefont{Bellwied, Borsanyi,
  Fodor, Katz, Pasztor, Ratti, and Szabo}}]{Bellwied:2015lba}
\bibinfo{author}{\bibfnamefont{R.}~\bibnamefont{Bellwied}},
  \bibinfo{author}{\bibfnamefont{S.}~\bibnamefont{Borsanyi}},
  \bibinfo{author}{\bibfnamefont{Z.}~\bibnamefont{Fodor}},
  \bibinfo{author}{\bibfnamefont{S.~D.} \bibnamefont{Katz}},
  \bibinfo{author}{\bibfnamefont{A.}~\bibnamefont{Pasztor}},
  \bibinfo{author}{\bibfnamefont{C.}~\bibnamefont{Ratti}}, \bibnamefont{and}
  \bibinfo{author}{\bibfnamefont{K.~K.} \bibnamefont{Szabo}},
  \bibinfo{journal}{Phys. Rev.} \textbf{\bibinfo{volume}{D92}},
  \bibinfo{pages}{114505} (\bibinfo{year}{2015}), \eprint{1507.04627}.

\bibitem[{\citenamefont{Borsanyi et~al.}(2012)\citenamefont{Borsanyi, Fodor,
  Katz, Krieg, Ratti, and Szabo}}]{Borsanyi:2011sw}
\bibinfo{author}{\bibfnamefont{S.}~\bibnamefont{Borsanyi}},
  \bibinfo{author}{\bibfnamefont{Z.}~\bibnamefont{Fodor}},
  \bibinfo{author}{\bibfnamefont{S.~D.} \bibnamefont{Katz}},
  \bibinfo{author}{\bibfnamefont{S.}~\bibnamefont{Krieg}},
  \bibinfo{author}{\bibfnamefont{C.}~\bibnamefont{Ratti}}, \bibnamefont{and}
  \bibinfo{author}{\bibfnamefont{K.}~\bibnamefont{Szabo}},
  \bibinfo{journal}{JHEP} \textbf{\bibinfo{volume}{01}}, \bibinfo{pages}{138}
  (\bibinfo{year}{2012}), \eprint{1112.4416}.

\bibitem[{\citenamefont{Bazavov et~al.}(2012)}]{Bazavov:2012jq}
\bibinfo{author}{\bibfnamefont{A.}~\bibnamefont{Bazavov}} \bibnamefont{et~al.}
  (\bibinfo{collaboration}{HotQCD}), \bibinfo{journal}{Phys. Rev.}
  \textbf{\bibinfo{volume}{D86}}, \bibinfo{pages}{034509}
  (\bibinfo{year}{2012}), \eprint{1203.0784}.

\bibitem[{\citenamefont{Pratt et~al.}(2015)\citenamefont{Pratt, Sangaline,
  Sorensen, and Wang}}]{Pratt:2015zsa}
\bibinfo{author}{\bibfnamefont{S.}~\bibnamefont{Pratt}},
  \bibinfo{author}{\bibfnamefont{E.}~\bibnamefont{Sangaline}},
  \bibinfo{author}{\bibfnamefont{P.}~\bibnamefont{Sorensen}}, \bibnamefont{and}
  \bibinfo{author}{\bibfnamefont{H.}~\bibnamefont{Wang}},
  \bibinfo{journal}{Phys. Rev. Lett.} \textbf{\bibinfo{volume}{114}},
  \bibinfo{pages}{202301} (\bibinfo{year}{2015}), \eprint{1501.04042}.

\bibitem[{\citenamefont{Denicol et~al.}(2010)\citenamefont{Denicol, Koide, and
  Rischke}}]{Denicol:2010xn}
\bibinfo{author}{\bibfnamefont{G.~S.} \bibnamefont{Denicol}},
  \bibinfo{author}{\bibfnamefont{T.}~\bibnamefont{Koide}}, \bibnamefont{and}
  \bibinfo{author}{\bibfnamefont{D.~H.} \bibnamefont{Rischke}},
  \bibinfo{journal}{Phys. Rev. Lett.} \textbf{\bibinfo{volume}{105}},
  \bibinfo{pages}{162501} (\bibinfo{year}{2010}), \eprint{1004.5013}.

\bibitem[{\citenamefont{Danielewicz and Pratt}(1996)}]{Danielewicz:1995ay}
\bibinfo{author}{\bibfnamefont{P.}~\bibnamefont{Danielewicz}} \bibnamefont{and}
  \bibinfo{author}{\bibfnamefont{S.}~\bibnamefont{Pratt}},
  \bibinfo{journal}{Phys. Rev.} \textbf{\bibinfo{volume}{C53}},
  \bibinfo{pages}{249} (\bibinfo{year}{1996}), \eprint{nucl-th/9507002}.

\end{thebibliography}

\end{document}